\documentclass[aps,prl,twocolumn,showpacs,superscriptaddress,groupedaddress]{revtex4}  
\usepackage{graphicx}  
\usepackage{dcolumn}   
\usepackage{bm}        
\usepackage{amssymb}   

\newcommand{\met}{\ensuremath{{\slash\kern-.7emE}_{T}}}
\newcommand{\metsub}{\ensuremath{{\slash\kern-.5emE}_{T}}}
\newcommand{\vmet}{\ensuremath{\vec{\slash\kern-.7emE}_{T}}}

\newcommand{\mt}{\ensuremath{m_T}}

\newcommand{\vut}{\ensuremath{\vec{u}_T}}

\newcommand{\wen}{\ensuremath{W \rightarrow e \nu}}

\newcommand{\pte}{\ensuremath{p_T^e}}

\newcommand{\ptnu}{\ensuremath{p_T^\nu}}

\newcounter{appendix}
\def\theappendix{\Alph{appendix}}
\def\appendix#1{
  \clearpage
  \addtocounter{appendix}{1}
  \addcontentsline{toc}{section}{\numberline{\theappendix}{#1}}
  \noindent{\bf\large Appendix \theappendix: #1}}

\newcounter {subsubsubsection}[subsubsection]

\def\pmcs{{\sc fastmc}}

\begin{document}

\hspace{5.2in} \mbox{FERMILAB-PUB-12-056-PPD}

\title{Measurement of the $\boldsymbol{W}$ Boson Mass with the D0 Detector}
%
\affiliation{LAFEX, Centro Brasileiro de Pesquisas F\'{i}sicas, Rio de Janeiro, Brazil}
\affiliation{Universidade do Estado do Rio de Janeiro, Rio de Janeiro, Brazil}
\affiliation{Universidade Federal do ABC, Santo Andr\'e, Brazil}
\affiliation{University of Science and Technology of China, Hefei, People's Republic of China}
\affiliation{Universidad de los Andes, Bogot\'a, Colombia}
\affiliation{Charles University, Faculty of Mathematics and Physics, Center for Particle Physics, Prague, Czech Republic}
\affiliation{Czech Technical University in Prague, Prague, Czech Republic}
\affiliation{Center for Particle Physics, Institute of Physics, Academy of Sciences of the Czech Republic, Prague, Czech Republic}
\affiliation{Universidad San Francisco de Quito, Quito, Ecuador}
\affiliation{LPC, Universit\'e Blaise Pascal, CNRS/IN2P3, Clermont, France}
\affiliation{LPSC, Universit\'e Joseph Fourier Grenoble 1, CNRS/IN2P3, Institut National Polytechnique de Grenoble, Grenoble, France}
\affiliation{CPPM, Aix-Marseille Universit\'e, CNRS/IN2P3, Marseille, France}
\affiliation{LAL, Universit\'e Paris-Sud, CNRS/IN2P3, Orsay, France}
\affiliation{LPNHE, Universit\'es Paris VI and VII, CNRS/IN2P3, Paris, France}
\affiliation{CEA, Irfu, SPP, Saclay, France}
\affiliation{IPHC, Universit\'e de Strasbourg, CNRS/IN2P3, Strasbourg, France}
\affiliation{IPNL, Universit\'e Lyon 1, CNRS/IN2P3, Villeurbanne, France and Universit\'e de Lyon, Lyon, France}
\affiliation{III. Physikalisches Institut A, RWTH Aachen University, Aachen, Germany}
\affiliation{Physikalisches Institut, Universit\"at Freiburg, Freiburg, Germany}
\affiliation{II. Physikalisches Institut, Georg-August-Universit\"at G\"ottingen, G\"ottingen, Germany}
\affiliation{Institut f\"ur Physik, Universit\"at Mainz, Mainz, Germany}
\affiliation{Ludwig-Maximilians-Universit\"at M\"unchen, M\"unchen, Germany}
\affiliation{Fachbereich Physik, Bergische Universit\"at Wuppertal, Wuppertal, Germany}
\affiliation{Panjab University, Chandigarh, India}
\affiliation{Delhi University, Delhi, India}
\affiliation{Tata Institute of Fundamental Research, Mumbai, India}
\affiliation{University College Dublin, Dublin, Ireland}
\affiliation{Korea Detector Laboratory, Korea University, Seoul, Korea}
\affiliation{CINVESTAV, Mexico City, Mexico}
\affiliation{Nikhef, Science Park, Amsterdam, the Netherlands}
\affiliation{Radboud University Nijmegen, Nijmegen, the Netherlands}
\affiliation{Joint Institute for Nuclear Research, Dubna, Russia}
\affiliation{Institute for Theoretical and Experimental Physics, Moscow, Russia}
\affiliation{Moscow State University, Moscow, Russia}
\affiliation{Institute for High Energy Physics, Protvino, Russia}
\affiliation{Petersburg Nuclear Physics Institute, St. Petersburg, Russia}
\affiliation{Instituci\'{o} Catalana de Recerca i Estudis Avan\c{c}ats (ICREA) and Institut de F\'{i}sica d'Altes Energies (IFAE), Barcelona, Spain}
\affiliation{Stockholm University, Stockholm and Uppsala University, Uppsala, Sweden}
\affiliation{Lancaster University, Lancaster LA1 4YB, United Kingdom}
\affiliation{Imperial College London, London SW7 2AZ, United Kingdom}
\affiliation{The University of Manchester, Manchester M13 9PL, United Kingdom}
\affiliation{University of Arizona, Tucson, Arizona 85721, USA}
\affiliation{University of California Riverside, Riverside, California 92521, USA}
\affiliation{Florida State University, Tallahassee, Florida 32306, USA}
\affiliation{Fermi National Accelerator Laboratory, Batavia, Illinois 60510, USA}
\affiliation{University of Illinois at Chicago, Chicago, Illinois 60607, USA}
\affiliation{Northern Illinois University, DeKalb, Illinois 60115, USA}
\affiliation{Northwestern University, Evanston, Illinois 60208, USA}
\affiliation{Indiana University, Bloomington, Indiana 47405, USA}
\affiliation{Purdue University Calumet, Hammond, Indiana 46323, USA}
\affiliation{University of Notre Dame, Notre Dame, Indiana 46556, USA}
\affiliation{Iowa State University, Ames, Iowa 50011, USA}
\affiliation{University of Kansas, Lawrence, Kansas 66045, USA}
\affiliation{Kansas State University, Manhattan, Kansas 66506, USA}
\affiliation{Louisiana Tech University, Ruston, Louisiana 71272, USA}
\affiliation{Boston University, Boston, Massachusetts 02215, USA}
\affiliation{Northeastern University, Boston, Massachusetts 02115, USA}
\affiliation{University of Michigan, Ann Arbor, Michigan 48109, USA}
\affiliation{Michigan State University, East Lansing, Michigan 48824, USA}
\affiliation{University of Mississippi, University, Mississippi 38677, USA}
\affiliation{University of Nebraska, Lincoln, Nebraska 68588, USA}
\affiliation{Rutgers University, Piscataway, New Jersey 08855, USA}
\affiliation{Princeton University, Princeton, New Jersey 08544, USA}
\affiliation{State University of New York, Buffalo, New York 14260, USA}
\affiliation{Columbia University, New York, New York 10027, USA}
\affiliation{University of Rochester, Rochester, New York 14627, USA}
\affiliation{State University of New York, Stony Brook, New York 11794, USA}
\affiliation{Brookhaven National Laboratory, Upton, New York 11973, USA}
\affiliation{Langston University, Langston, Oklahoma 73050, USA}
\affiliation{University of Oklahoma, Norman, Oklahoma 73019, USA}
\affiliation{Oklahoma State University, Stillwater, Oklahoma 74078, USA}
\affiliation{Brown University, Providence, Rhode Island 02912, USA}
\affiliation{University of Texas, Arlington, Texas 76019, USA}
\affiliation{Southern Methodist University, Dallas, Texas 75275, USA}
\affiliation{Rice University, Houston, Texas 77005, USA}
\affiliation{University of Virginia, Charlottesville, Virginia 22901, USA}
\affiliation{University of Washington, Seattle, Washington 98195, USA}
\author{V.M.~Abazov} \affiliation{Joint Institute for Nuclear Research, Dubna, Russia}
\author{B.~Abbott} \affiliation{University of Oklahoma, Norman, Oklahoma 73019, USA}
\author{B.S.~Acharya} \affiliation{Tata Institute of Fundamental Research, Mumbai, India}
\author{M.~Adams} \affiliation{University of Illinois at Chicago, Chicago, Illinois 60607, USA}
\author{T.~Adams} \affiliation{Florida State University, Tallahassee, Florida 32306, USA}
\author{G.D.~Alexeev} \affiliation{Joint Institute for Nuclear Research, Dubna, Russia}
\author{G.~Alkhazov} \affiliation{Petersburg Nuclear Physics Institute, St. Petersburg, Russia}
\author{A.~Alton$^{a}$} \affiliation{University of Michigan, Ann Arbor, Michigan 48109, USA}
\author{G.~Alverson} \affiliation{Northeastern University, Boston, Massachusetts 02115, USA}
\author{M.~Aoki} \affiliation{Fermi National Accelerator Laboratory, Batavia, Illinois 60510, USA}
% ...
%
%
\collaboration{The D0 Collaboration\footnote{with visitors from
$^{a}$Augustana College, Sioux Falls, SD, USA,
$^{b}$The University of Liverpool, Liverpool, UK,
$^{c}$UPIITA-IPN, Mexico City, Mexico,
$^{d}$DESY, Hamburg, Germany,
,
$^{e}$SLAC, Menlo Park, CA, USA,
$^{f}$University College London, London, UK,
$^{g}$Centro de Investigacion en Computacion - IPN, Mexico City, Mexico,
$^{h}$Thomas Jefferson National Accelerator Facility (JLab), Newport News, VA, USA,
$^{i}$ECFM, Universidad Autonoma de Sinaloa, Culiac\'an, Mexico,
$^{j}$Universidade Estadual Paulista, S\~ao Paulo, Brazil,
and
$^{k}$School of Physics, University of the Witwatersrand, Johannesburg, South Africa.
$^{\ddag}$Deceased.
}} \noaffiliation
\vskip 0.25cm

\date{March 1, 2012}

\begin{abstract}
We present a measurement of the $W$ boson mass using data corresponding to $4.3\ \mathrm{fb}^{-1}$ 
of integrated luminosity collected with the 
D0 detector during Run~II at the
Fermilab Tevatron $p\bar{p}$ collider. 
With a sample of 
1\,677\,394
$W\rightarrow e \nu$ candidate events, we measure 
$M_W = 80.367 \pm 0.026$~GeV.
This result is combined with an earlier D0 result determined using an independent
Run~II data sample, corresponding to $1\ \mathrm{fb}^{-1}$ of integrated luminosity, to yield
$M_W = 80.375 \pm 0.023\ \text{GeV}$.
\end{abstract}

\maketitle

In the context of the standard model (SM), there is a relationship between the $W$ boson mass ($M_W$) and the hypothetical Higgs boson mass (and other observables such as the top quark mass).  Accurate measurement of the $M_W$ is thus a key ingredient in constraining the SM Higgs boson mass and comparing that constraint with the results of direct Higgs boson searches. 
Precise measurements of $M_W$ have been reported by the ALEPH~\cite{AlephW}, DELPHI~\cite{DelphiW}, L3~\cite{L3W}, OPAL~\cite{OpalW}, D0~\cite{D0W,D0NewW}, and CDF~\cite{CDFW,CDFNewW} collaborations. 
The $W$ boson mass experimental methods and measurements are discussed in Ref.~\cite{ashujan}.
The current world average measured value is $M_W = 80.399 \pm 0.023$~GeV~\cite{b:mwwa}. 
This result and the current measurement~\cite{topmass} of the top quark mass, $M_t$, give a range for the predicted $M_H$ which is centered on a value outside the direct search allowed range. The predicted range is, however, large and does have some overlap with the regions allowed by direct searches. The limiting factor in the predictions is the experimental precision on $M_W$.
It is therefore of great interest to improve the precision of the $W$ boson mass measurement so as to further probe the validity of the SM.

In this Letter, we present a measurement of $M_W$ using data collected from
2006 to 2009 with the D0 detector~\cite{d0det}, corresponding to a total
integrated luminosity of 4.3~fb$^{-1}$.
We use the \wen\ decay mode because the D0 calorimeter is
well-suited for a precise measurement of electron~\cite{epm}
energies.
For the data considered in this analysis,
the average energy resolution 
is 4.2\% for electrons of 45 GeV. 
The longitudinal components of the colliding partons
and of the neutrino cannot be determined, so $M_W$ is determined using three kinematic
variables measured in the plane perpendicular to the beam direction:
the transverse mass \mt, the electron transverse momentum $\pte$, and the
neutrino transverse momentum $\ptnu$.  The transverse mass is defined as
$
\mt = \sqrt{2\pte \ptnu(1-\cos\Delta\phi)}, \label{eqn:mt}
$
where $\Delta\phi$ is the opening angle between the electron and neutrino
momenta in the plane transverse to the beam.  The vector $\vec{p}_T^{~\nu}$ is equal to 
the event missing transverse 
momentum
 ($\vmet$).

The D0 detector~\cite{d0det} 
comprises a tracking system, calorimeters and a muon system with an iron toroid magnet.
Silicon microstrip tracking detectors (SMT) near the interaction point 
cover $|\eta| < 3$, where $\eta \equiv - \ln [\tan( \theta/2)]$ and $\theta$  is the polar angle with respect to the proton beam direction, to provide tracking and vertex information. 
The central fiber tracker surrounds the SMT, providing coverage to 
$| \eta | \approx 2$.
A 1.9 T solenoid surrounds these tracking detectors.
Three uranium liquid-argon calorimeters measure particle energies.
The central calorimeter (CC) covers $| \eta | < 1.1$, and two end calorimeters
(EC) extend coverage to $| \eta | \approx 4$.  
The CC is segmented in depth into eight layers.  
The first four layers allow for a precise measurement of the energy of photons and electrons. 
The remaining four layers, along with the first four, are
used to measure the energy of hadrons.  
A three-level trigger system selects events for recording with a rate of $\approx$100~Hz.

The present analysis builds on the techniques developed in Ref.~\cite{D0NewW}. Additional studies are
necessary to cope with the consequences of the increased
instantaneous luminosities (on average $1.2 \times 10^{32} \ {\rm cm}^{-−2}{\rm s}^{-−1}$, almost 3
times higher than in Ref.~\cite{D0NewW}).
The main developments include a new 
model of dependence of the gains of the D0 calorimeter on the instantaneous luminosity.
This dependence had been predicted~\cite{TeV33} before the start of Run~II
and has been studied in detail in the data used for this Letter. 
The other important additions are a correction for residual $\eta$-dependent miscalibrations of the calorimeter response, a more detailed model of the impact of additional $p \overline p$ interactions on the electron energy reconstruction, and a detailed description of electron efficiency in the presence of additional $p \overline p$ interactions.
Using the same method as Ref.~\cite{D0NewW} we obtain the amount of material preceding the calorimeter from a fit to the longitudinal energy profile in the electromagnetic calorimeter.

Events are selected using a trigger requiring at least one electromagnetic (EM) cluster
found in the CC with the transverse energy threshold varying from 25 to 27~GeV
depending on run conditions.  The offline selection of candidate $W$~boson events
is similar to that used in Ref.~\cite{D0NewW}, except that the veto on electrons in $\phi$~regions
with degraded energy response is now based on extrapolation of the track
to the third calorimeter layer instead of the position of the calorimeter cluster.
We require at least one candidate electron reconstructed as an EM cluster
in the CC, matched in $(\eta,\phi)$ space to a track including at least one 
SMT hit and $p_T>10$~GeV to reject 
jets misidentified as electrons
and to ensure a precise measurement of the electron direction. 
The length of the electron three-momentum vector is defined by the cluster energy, and the direction by the track.
We require an electron with $\pte>25$~GeV that  passes shower shape and isolation requirements and points to the central 80\% in azimuth of a CC 
($|\eta|<1.05$) module.
The event must satisfy $~\met > 25$~GeV, $u_T < 15$~GeV, and $50 < \mt < 200$~GeV.  
Here $u_T$ is the magnitude of the vector sum of the transverse component of the 
energies measured in calorimeter cells excluding those associated with 
the reconstructed electron.
The relation $~\vmet=-(\vec{p}_T^{~e}+\vut)$ defines the missing momentum ascribed to the neutrino.
This selection yields 1\,677\,394 candidate $W\to e\nu$ events.

Candidate $Z\to ee$  events are required to have two EM clusters satisfying the above requirements, except
that one of the two may be reconstructed within an EC $(1.5<|\eta|<2.5)$.
The associated tracks must be of opposite curvature.  Events must also have $u_T < 15$~GeV and
$70 \le m_{ee} \le 110$~GeV, where $m_{ee}$ is the invariant mass of 
the electron pair.  
Events with both electrons in the CC are used to determine the calibration of the
electron energy scale.
There are 54\,512 candidate $Z\to ee$ events in this category. 
Events with one electron in EC are only used for the efficiency measurement.

The backgrounds in the $W$ boson candidate sample are $Z\to ee$ events 
where one electron escapes detection, multijet
events where a jet is misidentified as an electron with~~$\met$ arising
from misreconstruction, and $W\to \tau\nu \to e\nu\nu\nu$ events. 
The backgrounds are estimated using refined versions of 
the techniques in~Ref.~\cite{D0NewW}, and their impact on the 
measurement of $M_W$ is small. 
The fractions of the backgrounds in the $W$ boson candidate sample are
 $1.08$\%  for $Z\to ee$, $1.02$\% for multijet events, and $1.67$\% for $W\to\tau\nu\to e\nu\nu\nu$.

The {\sc resbos}~\cite{resbos} event generator, combined with {\sc photos}~\cite{photos}
is used to simulate the kinematics of $W$ and $Z$ boson production and decay.
{\sc resbos}  is a next-to-leading order event generator including next-to-next-to-leading logarithm resummation of soft gluons~\cite{resum}, and {\sc photos} generates up to two final state radiation photons.
Parton distribution functions (PDF) are described using CTEQ6.6~\cite{cteq66}.
This combination provides a good description of the most important effects
in the $M_W$ measurement, namely the boson transverse momentum spectrum
(influenced by the emission of multiple soft gluons) and radiation from the electrons
in the final state. We use comparisons to the {\sc wgrad}~\cite{wgrad} and {\sc zgrad}~\cite{zgrad}
event generators, which provide a more complete treatment of electroweak corrections at the one radiated photon level, 
in order to assess the uncertainty in the $M_W$ measurement due to quantum electrodynamics (QED) corrections.
We take the nonperturbative parameter $g_2$~\cite{g2} to be $0.68 \pm 0.02$~GeV$^2$~\cite{d0g2} and the uncertainty on $g_2$ is propagated to the $W$ boson mass uncertainty.

A fast, parametrized Monte Carlo (MC) simulation (\pmcs) is used to simulate electron identification efficiencies and the energy response and resolutions of the electron and recoil system in the generated events.
The \pmcs\ parameters are determined using a combination of detailed simulation and
control data samples. 
The primary control sample used for both the
electromagnetic and hadronic response tuning is $Z\to ee$ events. Events recorded 
in random beam crossings are overlaid on $W$ and $Z$ events in the detailed
simulation to quantify the effect of additional collisions in the same or nearby bunch crossings.

The $Z$ boson mass and width are known with high precision from
measurements at LEP~\cite{ZLEP}. These values are used to calibrate the
electromagnetic calorimeter response assuming a form $E^{\text{meas}} =
\alpha\,E^{\text{true}} + \beta$ with constants $\alpha$ and $\beta$ determined
from fits to the dielectron mass spectrum and the energy and angular distributions of the two electrons.
The $M_W$ measurement presented here is
effectively a measurement of the ratio of $W$ and $Z$ boson masses.

The hadronic energy in the event contains the hadronic system recoiling from the $W$ boson, the effects of low energy products from spectator parton collisions and other beam collisions, final state radiation, and energy from the recoil particles that enter the electron selection window.
The hadronic response (resolution) is calibrated using the mean (width) of the $\eta_{\text{imb}}$ distribution in $Z\to ee$ events in bins of $p_T^{ee}$. 
Here, $\eta_{\text{imb}}$ is defined as the projections of the sum
of dielectron transverse momentum $(\vec{p}_T^{~ee})$ and $\vec{u}_T$ vectors on the axis
bisecting the dielectron directions in the transverse plane~\cite{ua2eta}.

The combination of event generator and \pmcs\ is used to predict the shapes
of $\mt$, $\pte$, and $\met$ for a given $M_W$ hypothesis. 
$M_W$ is determined separately for each of the three observables by maximizing a
binned likelihood between the data distribution and the predicted distribution normalized
to the data.
The fit ranges are optimized as indicated in Table~\ref{tab:fits}.

A test of the analysis procedure is performed using 
$W\to e\nu$ events, generated by the {\sc pythia}~\cite{pythia} event generator and processed through a
detailed {\sc geant} MC simulation~\cite{b:geant}, 
which are treated as collider data.  
The {\sc fastmc} is separately tuned to give agreement with the {\sc geant} events in the same way as for the data comparison.
Each of the $M_W$ fit results using the $\mt$, $\pte$, and
$\met$ distributions agree with the input $M_W$ value within the
6~MeV total uncertainty of the test arising from MC statistics. 

\begin{figure}[b]
  \includegraphics[width=0.96\linewidth]{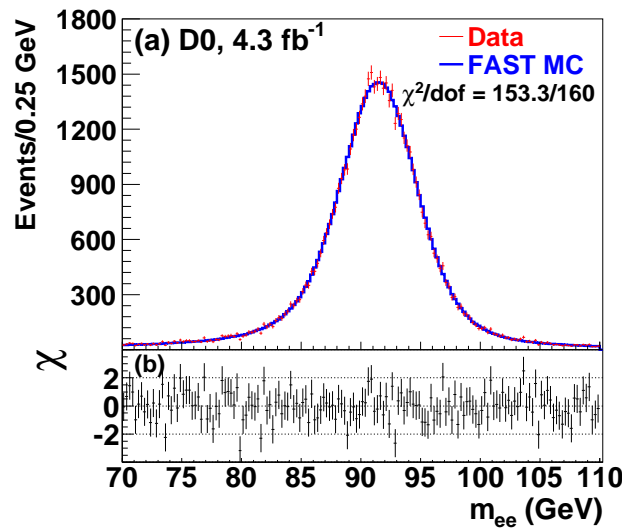}
  \caption{(a) The dielectron invariant mass distribution in $Z\to ee$ data and 
    from the \pmcs\ and (b) the $\chi$ values, 
    where $\chi_i = [N_i-\,($\pmcs$_i)]/\sigma_i$ for each bin in the distribution, 
    $N_i$ and \pmcs$_i$ are the data and \pmcs\ template yields in bin $i$, respectively,
    and $\sigma_i$ is the statistical uncertainty in bin $i$.
    \label{f:zfinal}}
\end{figure}

\begin{figure*}[hbpt]
  \includegraphics[width=0.32\linewidth]{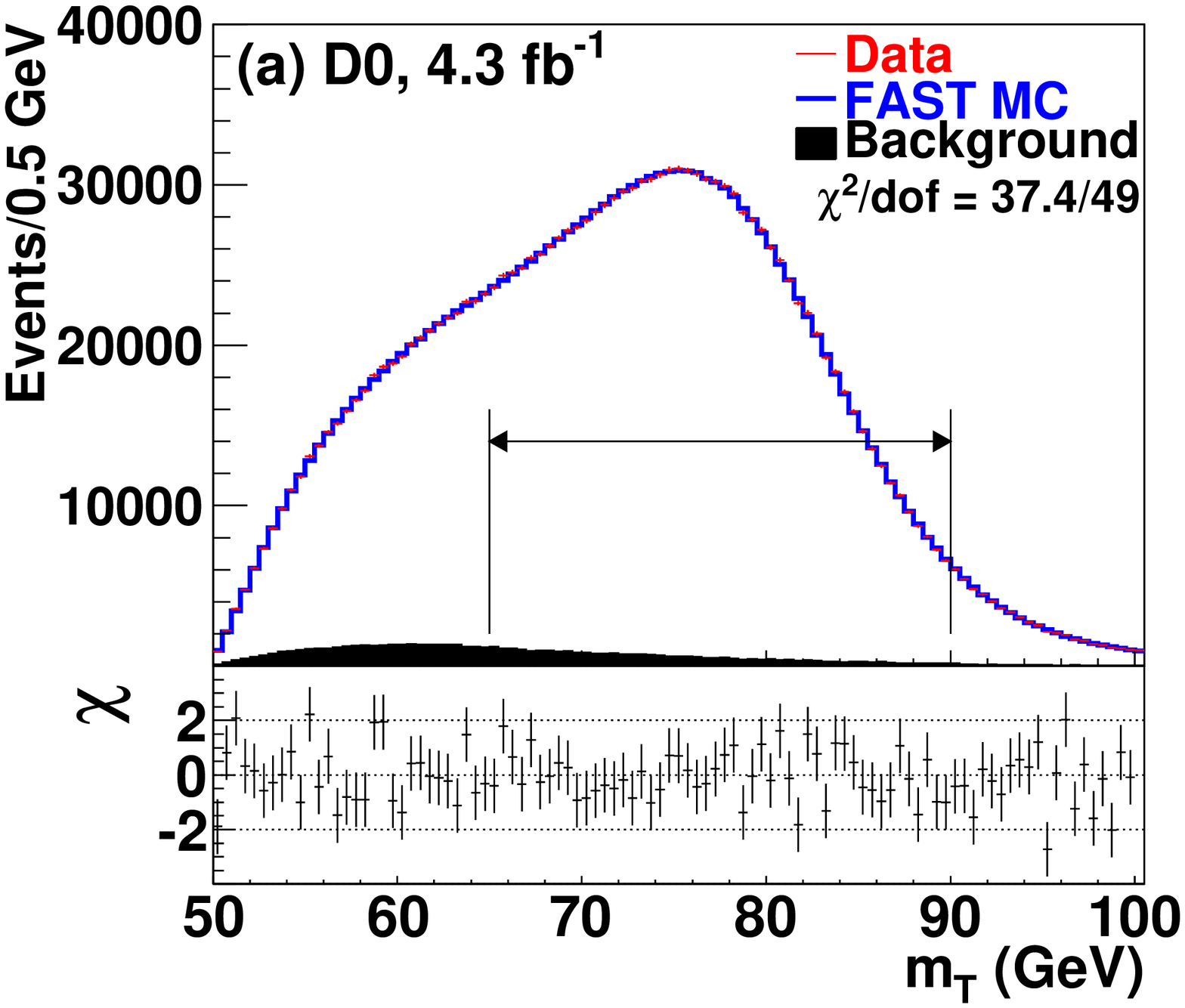}
  \includegraphics[width=0.32\linewidth]{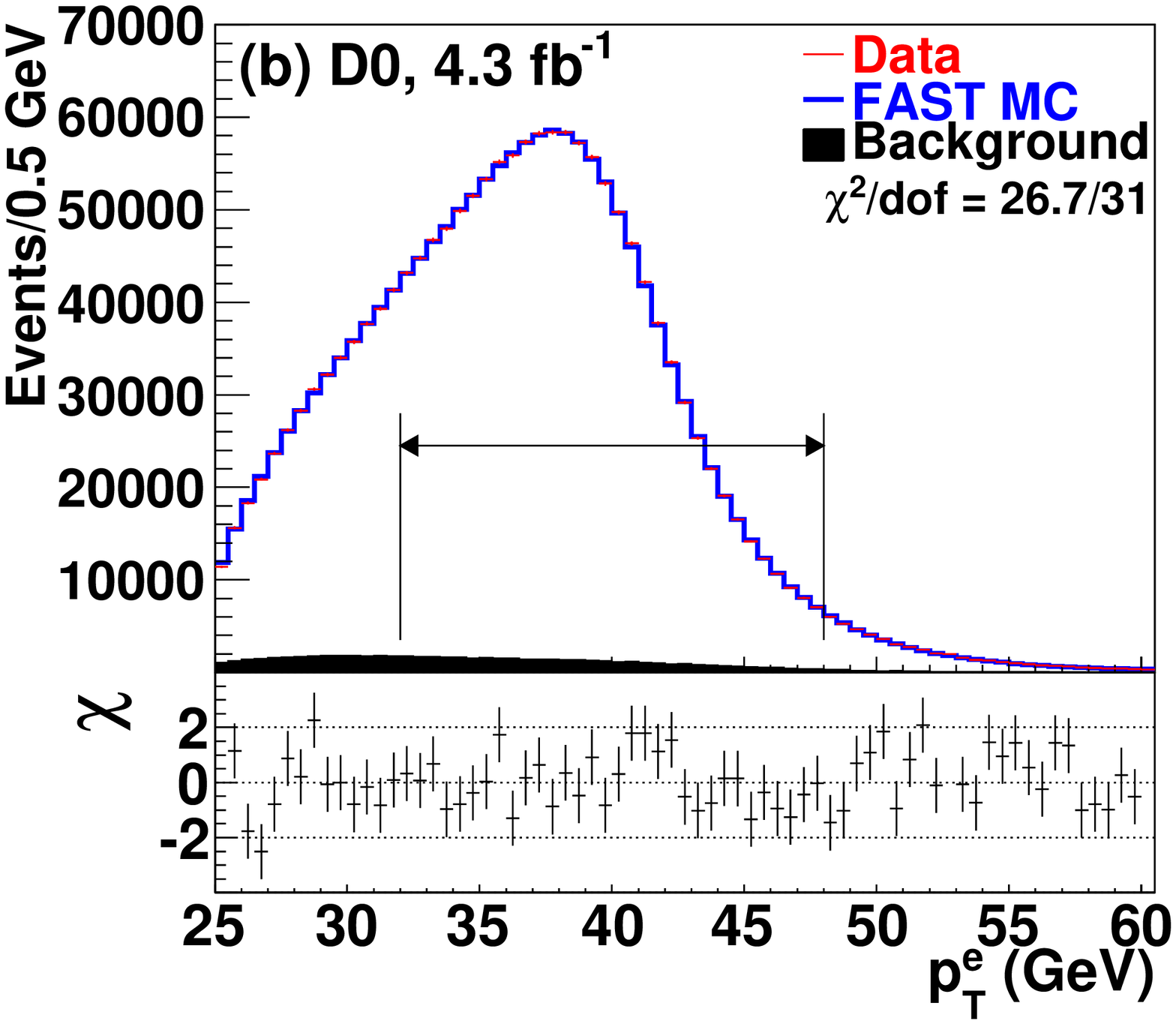}
  \includegraphics[width=0.32\linewidth]{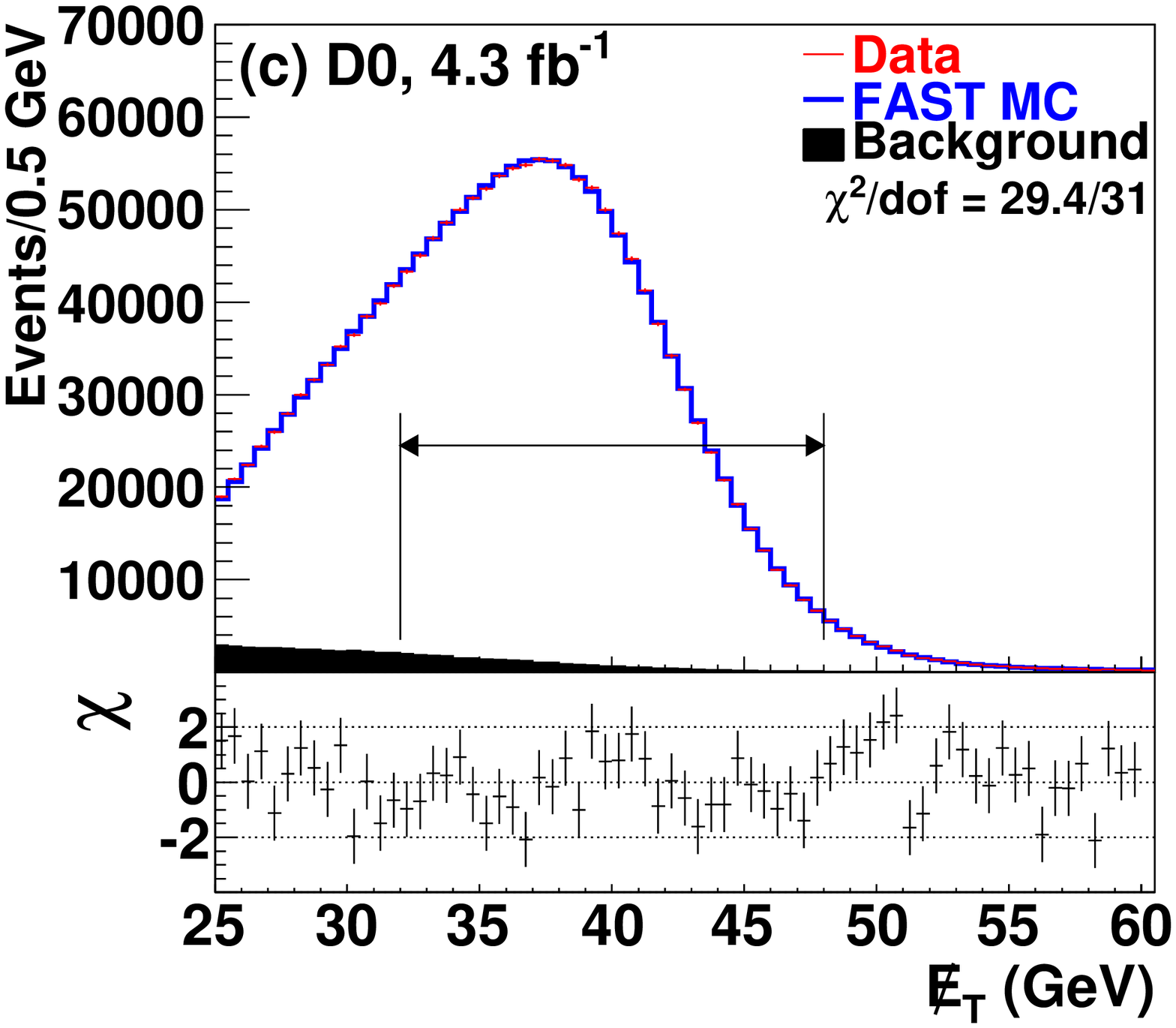}
  \caption{The (a) $\mt$, (b) $\pte$, and (c) $\met$ distributions for data and
    \pmcs\ simulation with backgrounds. The $\chi$ values are shown below each
    distribution, where $\chi_i = [N_i-\,($\pmcs$_i)]/\sigma_i$ for each bin in
    the distribution,   
    $N_i$ and \pmcs$_i$ are the data and \pmcs\ template yields in bin $i$, respectively,
    and $\sigma_i$ is the statistical uncertainty in bin $i$.
    The fit  ranges are indicated by the double-ended horizontal arrows.\label{fig:fits}}
\end{figure*}

During the \pmcs\ tuning performed to describe the collider data, the $M_W$ values returned
from fits had an unknown constant offset added.  The same
offset was used for $\mt$, $\pte$ and $\met$.  This
allowed the full tuning on the $W$ and $Z$ boson events and internal consistency
checks to be performed without knowledge of the final result.
Once the important data and \pmcs\ comparison plots had acceptable $\chi^2$
distributions, 
the common offset was removed from the results.
The $Z$ boson mass from the fit to the data corresponds to the input that was used in the determination of the calorimeter response described above. The statistical uncertainty from the fit is 0.017~GeV, quoted here as a quantitative illustration of the statistical power of the $Z\to ee$ sample.
Figure~\ref{f:zfinal} shows a comparison of the $m_{ee}$ distributions for data
and \pmcs.
The $M_W$ results
are given in
Table~\ref{tab:fits}.  The $\mt$, $\pte$, and $\met$ distributions showing the data and \pmcs\ templates
with background for the best fit $M_W$ are shown in Fig.~\ref{fig:fits}.

\begin{table}[hbtp]
\begin{center}
  \caption{Results from the fits to data.  
  The uncertainty is solely due to the statistics of the $W$ boson sample.
  \label{tab:fits}}
  \begin{tabular}{cccc}
     \hline\hline
     Variable & Fit Range (GeV) & $M_W$ (GeV)     & $\ \ \ \chi^2$/dof \\ \hline
      $\mt$   & $65<\mt<90$      & $\ \ \ 80.371\pm0.013\ \ \ $ &  37.4/49  \\
     $\pte$   & $32<\pte<48$      & $      80.343\pm0.014      $ &  26.7/31  \\ 
     $\met$   & $32<\met<48$    & $      80.355\pm0.015      $ &  29.4/31  \\
     \hline\hline
  \end{tabular}
\end{center}
\end{table}

The systematic uncertainties in the $M_W$ measurement are summarized in 
Table~\ref{t:syst}. 
They can be categorized as those from experimental sources
and those from uncertainties in the production mechanism.
The uncertainties on the electron energy calibration, the electron energy resolution,
and the hadronic recoil model arise from the finite size of the $Z\to ee$ sample
used to derive them.
The uncertainties in the propagation of electron energy calibrations from the $Z\to ee$ to 
the $W\to e\nu$ sample are determined by the difference in energy loss in 
the uninstrumented material in front of the calorimeter.
The energy loss as a function of electron energy and~$\eta$ is derived
from a 
dedicated detailed {\sc geant} 
simulation of the D0 detector. 
The shower modeling systematic uncertainties reflect the uncertainties 
in the amount of uninstrumented material, and the energy loss systematic uncertainties arise
from the finite precision of our simulations of electron showers
based on a detailed model of the detector geometry.
The systematic uncertainties of electron efficiency, hadronic recoil model, and backgrounds are determined by varying the corresponding parameters within the statistical uncertainties of their measurements. 
Table~\ref{t:syst} also shows the $M_W$ uncertainties arising from the
backgrounds.

The uncertainties due to the production mechanism are dominated
by the uncertainties due to the PDFs. 
The transverse observables ($\mt$, $\pte$, and $\met$) used in the measurement of $M_W$ are invariant under longitudinal boosts, and their use therefore minimizes the sensitivity to PDF uncertainties. However, a limited sensitivity to PDF uncertainties does arise from the electron pseudorapidity requirements that are used in the measurement of $M_W$ reported here. These requirements are not invariant under longitudinal boosts, and changes in the PDF can therefore result in changes of the shapes of our transverse observables.
The uncertainties in the PDF are 
propagated to a 1 standard deviation uncertainty in $M_W$ by generating ensembles 
of $W$ boson events using {\sc pythia} with 
the CTEQ6.1~\cite{cteq61} prescription. The other production uncertainties have been discussed above.

\begin{table}
\begin{center}
  \caption{Systematic uncertainties of the $M_W$  measurement.\label{t:syst}}
  \begin{ruledtabular}
  \begin{tabular}{ lccc}
          & \multicolumn{3}{c}{$\Delta M_W$~(MeV)} \\
   Source                          &$\mt$ & $\pte$ &  $\met$\\
  \hline \hline
  Electron energy calibration       & 16 &  17 & 16 \\
  Electron resolution model         &  2 &   2 &  3 \\
  Electron shower modeling           &  4 &   6 &  7 \\
  Electron energy loss model        &  4 &   4 &  4 \\
  Hadronic recoil model             &  5 &  6 & 14 \\
  Electron efficiencies             &  1 &   3 &  5 \\
  Backgrounds                       &  2 &   2 &  2 \\ \hline
  Experimental subtotal             & 18 &  20 & 24 \\ \hline
				    				     
  PDF                          &  11 &  11 & 14 \\
  QED                          &  7 &   7 &  9 \\
  Boson $p_T$                  &  2 &   5 &  2 \\ \hline
  Production subtotal          & 13 &  14 & 17 \\ \hline

  Total                        &  22 & 24 & 29 \\
  \end{tabular}
  \end{ruledtabular}
\end{center}
\end{table}    

The quality of the simulation is indicated by the $\chi^2$
values computed for the differences
between the data and \pmcs\ shown in Figs.~\ref{f:zfinal} and~\ref{fig:fits}.  
We perform a variety of consistency checks of the stability of our results.  We vary the fit ranges for
the $\mt$, $\pte$ and $\met$ distributions.
The data are also divided
into statistically independent categories based on instantaneous luminosity,
time, electron~$\eta$, and the projection of \vut\ on the 
electron direction. 
The exclusion region near CC module edges is varied, and the selection requirement on\ $u_T$ is varied.
The results are stable to within 
the measurement uncertainty for each of these tests.

The total correlations among the three $W$ boson mass measurements are determined by combining the covariance matrices for each source of uncertainty.  For uncertainties which arise from sample statistics, such as the electron energy scale, the full covariance matrices are determined using ensemble studies.  For uncertainties which are nonstatistical in nature, such as the QED uncertainty, the correlations among the three observables are defined as 100\% to prevent these uncertainties from being decreased in the combination.  
The resulting total correlations, including both categories of uncertainties, are 0.89 $(\mt,\,\pte)$, 0.86 $(\mt,\,\met)$ and 0.75 $(\pte,\,\met)$.  When considering only the uncertainties which are allowed to decrease in the combination, we find that the $\met$ measurement has negligible weight.  
We therefore combine the $\mt$ and $\pte$ measurements using the method~\cite{b:BLUE} and obtain
\begin{eqnarray*}
  M_W & = & 80.367 \pm 0.013\ \mathrm{(stat.)} \pm 0.022\ \mathrm{(syst.)\ GeV}\\
      & = & 80.367 \pm 0.026\ \mathrm{GeV}.
\end{eqnarray*}

The probability to observe a larger difference than observed between these two measurements is 2.8\%. 
The probability to observe a larger difference than observed when all three measurements are combined is 5\%.  
We combine this measurement with the earlier D0 measurement~\cite{D0NewW} to obtain
\begin{eqnarray*}
  M_W & = & 80.375 \pm 0.011\ \mathrm{(stat.)} \pm 0.020\ \mathrm{(syst.)\ GeV}\\
      & = & 80.375 \pm 0.023\ \mathrm{GeV}.
\end{eqnarray*}

\begin{figure}[tb]
  \includegraphics[width=0.96\linewidth]{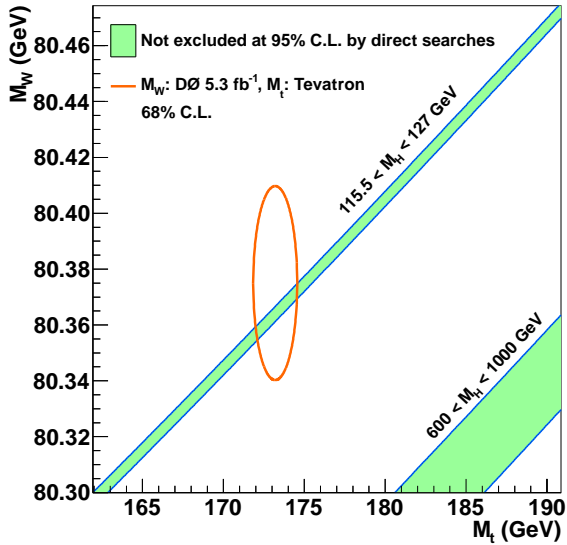}
  \caption{Contour curves of 68~\% probability in the ($M_t,M_W$) plane. The ellipse
represents the measurement of $M_t$ from Ref.~\cite{topmass} and the measurement of $M_W = 80.375 \pm 0.023\ \mathrm{GeV}$
reported in this Letter. The bands show the SM prediction for different Higgs boson mass hypotheses that are
not yet ruled out by direct searches~\cite{higgssearch} for the Higgs boson.
    \label{f:MMplane}}
\end{figure}

The dominant uncertainties arise from the available statistics of the $W\to
e\nu$ and $Z\to ee$ samples.  
Thus, a future measurement with the full D0 dataset is expected to be more precise.
The $M_W$ measurement reported here agrees with the world average~\cite{b:mwwa,combineshift}
and the previous individual measurements and 
has an uncertainty that significantly improves upon previous D0 measurements.
Our new measurement of $M_W$ and the most recent world average measurement of $M_t$ are compared in Fig.~\ref{f:MMplane} with the regions that are still allowed, at the 95\% C.L., after direct searches for the Higgs boson at LEP, the Tevatron and the LHC.
Our new measurement of $M_W$ is in good agreement with one of the regions allowed by direct searches for the Higgs boson.


%
We thank the staffs at Fermilab and collaborating institutions,
and acknowledge support from the
DOE and NSF (USA);
CEA, CNRS/IN2P3 and the CIMENT project, Grenoble (France);
MON, Rosatom and RFBR (Russia);
CNPq, FAPERJ, FAPESP and FUNDUNESP (Brazil);
DAE and DST (India);
Colciencias (Colombia);
CONACyT (Mexico);
NRF (Korea);
FOM (The Netherlands);
STFC and the Royal Society (United Kingdom);
MSMT and GACR (Czech Republic);
BMBF and DFG (Germany);
SFI (Ireland);
The Swedish Research Council (Sweden);
and
CAS and CNSF (China).
%


\begin{thebibliography}{99}

\bibitem{AlephW} S.~Schael {\sl et al.} (ALEPH Collaboration), Eur. Phys. J. 
   C {\bf 47}, 309 (2006).
   
\bibitem{DelphiW} J.~Abdallah {\sl et al.} (DELPHI Collaboration),
  Eur. Phys. J. C {\bf 55}, 1 (2008).
  
\bibitem{L3W} P.~Achard {\sl et al.} (L3 Collaboration), Eur. Phys. J. C {\bf 45}, 
  569 (2006).
  
\bibitem{OpalW} G.~Abbiendi {\sl et al.} (OPAL Collaboration), Eur. Phys. J. 
  C {\bf 45}, 307 (2005).
  
\bibitem{D0W} B.~Abbott {\sl et al.} (D0 Collaboration), Phys. Rev. D {\bf 58}, 092003 (1998); 
 B.~Abbott {\sl et al.} (D0 Collaboration), Phys. Rev. D {\bf 62}, 092006 (2000); 
 V.~M.~Abazov {\sl et al.} (D0 Collaboration), Phys. Rev. D {\bf 66}, 012001 (2002).
 
\bibitem{D0NewW} V.~M.~Abazov {\sl et al.} (D0 Collaboration), Phys. Rev. Lett. {\bf 103}, 141801 (2009).

\bibitem{CDFW} T.~Affolder {\sl et al.} (CDF Collaboration), Phys. Rev. D {\bf 64},
  052001 (2001).
  
\bibitem{CDFNewW} T.~Aaltonen {\sl et al.} (CDF Collaboration), Phys. Rev. Lett. 
 {\bf 99}, 151801 (2007); T.~Aaltonen {\sl et al.} (CDF Collaboration), Phys. Rev. D 
 {\bf 77}, 112001 (2008).
 

\bibitem{ashujan} A.~V.~Kotwal and J.~Stark, Ann. Rev. Nucl. Part. Sci. {\bf 58}, 147 (2008).



\bibitem{b:mwwa} The LEP Electroweak Working Group and the Tevatron Electroweak Working Group, CERN-PH-EP-2010-095, FERMILAB-TM-2480-PPD, arXiv:1012.2367 [hep-ex] (2010).

\bibitem{topmass} Tevatron Electroweak Working Group, CDF and D0 Collaborations,  arXiv:1107.5255 [hep-ex]  (2011).

\bibitem{d0det} V.~M.~Abazov {\sl et al.} (D0 Collaboration),
   Nucl. Instrum. Methods in Phys. Res. A {\bf 565}, 463  (2006).
   
\bibitem{epm} Throughout this Letter we use electron to imply either electron or positron.

\bibitem{TeV33} H.~Montgomery {\sl et al.} (D0 Collaboration),
arXiv:hep-ex/9804011 (1998).

\bibitem{resbos}
  C.~Balazs and C.~P.~Yuan, Phys. Rev. D {\bf 56}, 5558 (1997).
  
\bibitem{photos}
  P.~Golonka and Z.~Was, Eur. Phys. J. C {\bf 45}, 97 (2006).
  
\bibitem{resum} J.~C.~Collins, D.~E.~Soper, G.~Sterman, Nucl. Phys {\bf B} 250, 199 (1985).

\bibitem{cteq66}
P.~M.~Nadolsky, H.~-L.~Lai, Q.~-H.~Cao, J.~Huston, J.~Pumplin, D.~Stump, W.~-K.~Tung, and C.~-P.~Yuan, Phys. Rev. D {\bf 78}, 013004 (2008).

\bibitem{wgrad}
  U.~Baur, S.~Keller, and D.~Wackeroth, Phys. Rev. D {\bf 59}, 013002 (1998).
  
\bibitem{zgrad} 
  U.~Baur, S.~Keller, and W.~K.~Sakumoto, Phys. Rev. D {\bf57}, 199 (1998);
  U.~Baur, O.~Brein, W.~Hollik, C.~Schappacher, and D.~Wackeroth, Phys. Rev. D {\bf65}, 033007 (2002).

\bibitem{g2} 
  F.~Landry, R.~Brock, P.~M.~Nadolsky, and C.~P.~Yuan, Phys. Rev. D {\bf 67}, 
    073016 (2003).
    
\bibitem{d0g2}
  V.~M.~Abazov {\sl et al.}, (D0 Collaboration), Phys. Rev. Lett. {\bf 100}, 102002 (2008).

\bibitem{ZLEP} ALEPH Collaboration, DELPHI Collaboration, L3 Collaboration,
OPAL Collaboration, SLD Collaboration, LEP Electroweak
Working Group, and SLD Electroweak and Heavy Flavour Groups, Phys. Rept.
{\bf 427}, 257 (2006).


\bibitem{ua2eta} J.~Alitti {\sl et al.} (UA2 Collaboration), Phys. Lett. B {\bf 276}, 354 (1992).


\bibitem{pythia} T.~Sj\"{o}strand, S.~Mrenna, and P.~Skands, J. High Energy Phys. {\bf 05}, 026 (2006).

\bibitem{b:geant}  R.~Brun and F.~Carminati, CERN Program Library Long
  Writeup, Report No. W5013, (1993).



\bibitem{cteq61}  H.~L.~Lai,  J.~Huston, S.~Kuhlmann, F.~Olness, J.~Owens, D.~Soper, W.~K.~Tung, and H.~Weerts, Phys. Rev. D {\bf 55}, 1280 (1997); 
D.~Stump, J.~Huston, J.~Pumplin, W.~-K.~Tung, H.~-L.~Lai, S.~Kuhlmann, and J.~F.~Owens, J. High Energy Phys. {\bf 10}, 046 (2003).



\bibitem{b:BLUE} L.~Lyons, D.~Gibout, and P.~Clifford, Nucl. Instrum. Methods
    in Phys. Res. A {\bf 270}, 110 (1988); A.~Valassi, Nucl. Instrum. Methods
    in Phys. Res. A {\bf 500}, 391 (2003).
    
\bibitem{combineshift} RESBOS uses a $W$ width of 2100.4~MeV.  Combinations~\cite{b:mwwa} normally use a standard model width of $2093.2\pm2.2$~MeV (using the current world average $m_W$), which would require a correction of $1.1\pm0.5$~MeV to our quoted result.


\bibitem{higgssearch} G.~Abbiendi {\sl et al.} (ALEPH Collaboration, DELPHI Collaboration, L3 Collaboration, OPAL Collaboration, and LEP Working Group for Higgs Boson Searches), Phys. Lett. B {\bf 565}, 61 (2003);
Tevatron New Phenomena and Higgs Working Group, CDF, and D0 collaborations, arXiv:1203.3774 [hep-ex] (2012); 
G.~Aad {\sl et al.} (ATLAS Collaboration), Phys. Lett. B {\bf 710}, 49 (2012);
S.~Chatrchyan {\sl et al.} (CMS Collaboration), Phys. Lett. B {\bf 710}, 26 (2012).


\end{thebibliography}
\end{document}